\documentclass[aps,pra,twocolumn,showpacs,floatfix,superscriptaddress]{revtex4-2}
\usepackage{graphicx}
\usepackage{caption}
\usepackage[table]{xcolor}
\usepackage{amsmath}
\usepackage{amssymb}
\usepackage{amsfonts}
\usepackage{bm}
\usepackage{hyperref}
\usepackage{url}
\usepackage{multirow}
\usepackage{array}
\usepackage{cleveref}
 \usepackage{soul}
\usepackage[detect-none]{siunitx}
\usepackage[english]{babel}

\colorlet{bckgr}{white!100!}
\renewcommand{\st}[1]{\unskip}

\sethlcolor{bckgr}
\setstcolor{red}

\newsavebox{\tablebox}
\graphicspath{{./figures/}}
\newcommand{\la}{\langle}
\newcommand{\ra}{\rangle}

\newcommand{\Tm}{Tm\,{\sc i}}
\newcommand{\etal}{{\it et al.}}
\newcolumntype{F}[1]{%
    >{\raggedright\arraybackslash\hspace{0pt}}p{#1}}%
\newcolumntype{T}[1]{%
    >{\centering\arraybackslash\hspace{0pt}}p{#1}}%
\begin{document}
\thispagestyle{empty}
\title{Comparison of theory and experiment for radiative characteristics in neutral thulium}

\author{Andrey I. Bondarev}
\email{a.bondarev@hi-jena.gsi.de}
\affiliation{Helmholtz Institute Jena, 07743 Jena, Germany}
\affiliation{GSI Helmholtzzentrum f\"ur Schwerionenforschung GmbH, 64291 Darmstadt, Germany}
\author{Maris Tamanis}
\affiliation{Laser Centre, Faculty of Physics, Mathematics and Optometry, University of Latvia, LV-1586 Riga, Latvia}
\author{Ruvin Ferber}
\email{ruvins.ferbers@lu.lv}
\affiliation{Laser Centre, Faculty of Physics, Mathematics and Optometry, University of Latvia, LV-1586 Riga, Latvia}
\author{G\"on\"ul~Ba\c{s}ar}
\affiliation{Istanbul University, Faculty of Science, Physics Department, TR-34134 Vezneciler, Istanbul,  T\"urkiye }
\author{Sophie Kr\"oger}
\affiliation{Hochschule f\"ur Technik und Wirtschaft Berlin, 12459 Berlin, Germany}
\author{Mikhail G. Kozlov}
\affiliation{St. Petersburg Electrotechnical University ``LETI'', St. Petersburg 197376, Russia}
\author{Stephan Fritzsche}
\affiliation{Helmholtz Institute Jena, 07743 Jena, Germany}
\affiliation{GSI Helmholtzzentrum f\"ur Schwerionenforschung GmbH, 64291 Darmstadt, Germany}
\affiliation{Theoretisch-Physikalisches Institut, Friedrich-Schiller-Universit\"at Jena, 07743 Jena, Germany}
\date{\today}
\begin{abstract}
Intensities in the \Tm\ emission series originating from a common upper level are measured using a Fourier transform spectrometer. The derived relative transition probabilities within each series are compared to the theoretical predictions obtained from large-scale calculations that combine configuration interaction with many-body perturbation theory. Moreover, the \Tm\ spectrum recorded in an external magnetic field is analyzed.
Our theoretical results well describe the current measurements and show no more than a two-fold difference from previous experimental data on absolute transition probabilities. 
Additionally, Land\'e $g$ factors, hyperfine structure constants, and atomic electric quadrupole moments for several levels of interest are computed and compared to experimental observations, where available. 
\end{abstract}

%
\maketitle
\section{Introduction}

\hl{The paper is aimed to perform measurements of radiative characteristics of neutral thulium based on spectral intensities
and comparison to the modern theory.
} Thulium ($Z=69$) is the 13th element of the lanthanide series with a ground electron configuration of the neutral atom [Xe]$4f^{13} 6s^2$.
Experimental research devoted to the identification of energy levels and transitions in this atom has a long history, with Sugar~\etal~\cite{Sugar_73} classifying the spectrum.
%
Extensive investigations employing a variety of techniques were focused on the hyperfine structure~\cite{Ritter_62,Kuhl_71,Brandt_77,vLeeuwen_80,Childs_84, Kroeger_97, Basar_05, Fedorov_15, par_22, keb_22}. 
%
Radiative lifetime measurements began with Handrich~\etal~\cite{Handrich_69} and Wallenstein~\cite{Wallenstein_72}, who used the zero-field level crossing technique to study levels in the lowest excited configurations. Penkin and Komarovsky~\cite{Penkin_76} along with Blagoev~\etal~\cite{Blagoev_78} employed different methods, such as the method of hooks and delayed coincidence, to measure lifetimes of higher-lying levels. Some of these measurements were later confirmed with better accuracy by Zaki Ewiss~\etal~\cite{Zaki_84} using a pulsed dye laser. 
Later on, Anderson~\etal~\cite{Anderson_96} reported lifetime measurements based on time-resolved laser-induced fluorescence of a large number of levels. Using their results and analysing relative intensity distributions in Fourier Transform (FT) spectra, Wickliffe and  Lawler~\cite{wic_97} determined absolute transition probabilities and branching fractions for many levels. More recently Tian~\etal~\cite{Tian_16} extended the number of levels with known lifetimes. Additionally, Wang \etal~\cite{Wang_22} determined the branching fractions for numerous lines, utilizing their data and the same emission spectra from the library of the USA National Solar Observatory~\cite{wic_97}. It is worth noting that some of the referenced studies on lifetimes and branching fractions also investigated the singly ionized thulium ion, Tm~\textsc{ii}, alongside neutral \Tm. 

Theoretical investigations are relatively scarce. Camus~\cite{Camus_66} carried out early calculations of energy levels and Land\'e $g$ factors, as well as oscillator strengths and lifetimes~\cite{Camus_70}. Pfeufer~\cite{Pfeufer_86} performed semiempirical calculations of the hyperfine structure. Cheng and Childs~\cite{Cheng_85} used the multiconfiguration Dirac-Fock method to compute excitation energies, $g$ factors, and hyperfine structure constants for the ground state multiplet of thulium along with other rare-earth atoms. More recently, Li and Dzuba~\cite{li_20} calculated energy levels and electric dipole transition amplitudes between ground and low-lying states as well as ionization potentials, and electron aﬃnities for thulium and its heavier homologue, mendelevium.

The magnetic dipole transition at 1.14~\unit{\micro\metre} between the fine-structure levels within the ground electronic configuration of \Tm\ has been proposed and validated as a promising candidate for \hl{operating optical clocks with high precision}\st{optical clock operation at low uncertainty}~\cite{Sukachev_16,GFTS19,Golovizin_21,GTYM23}. Gaire~\etal~\cite{Gaire_19,Gaire_23} investigated the clock transition in thulium atoms trapped in solid noble gas crystals. The experimental advancements led to increased interest in thulium clock states from a theoretical standpoint. Recently, Dzuba~\cite{Dzuba_20} calculated
polarizabilities of these states and Fleig~\cite{fle_23}, their electric quadrupole moments.

Atoms with an open $f$-shell have very complex and dense spectra. Such systems can be very sensitive to small perturbations, which makes them important tools for exploring physics beyond the Standard model \cite{Safronova_18}. Additionally, some of these atoms exhibit elements of quantum chaos \cite{FGGK94,VKF17}. Furthermore, accurate predictions of the probabilities of \st{both} strong permitted electric dipole as well as weaker forbidden transitions in elements with complex electronic structure are of current interest in astrophysics~\cite{Tanaka_20,Bondarev_23}. However, the theoretical treatment of these atoms is very difficult. The $f$-electrons are located close to the nucleus, and when the occupation number of the $f$-shell changes, it results in a drastic change in the mean field experienced by other valence electrons. Accounting for this effect requires long basis sets and very large configuration spaces in calculations; see also Ref.~\cite{Fritzsche_22}.

In this study, we analyze the FT spectrum of neutral thulium within the visible spectral range. This spectrum was previously utilized in Ref.~\cite{par_22} for a hyperfine structure investigation.  From this spectrum we determine the relative intensity distributions in branches \st{originating} \hl{that originate} from a common upper level. In our current investigation we identify new branches that were not reported in Refs.~\cite{wic_97,Wang_22}. Additionally, we have recorded new spectra in an external magnetic field and present the Zeeman splitting of an exemplary line.
Our experimental findings are \st{also} compared to theoretical outcomes derived using a method that combines the configuration interaction with many-body perturbation theory~\cite{dzu_96,koz_22}. As a result of these large-scale calculations, we obtain predictions for \hl{level energies,} Land\'e $g$ factors, hyperfine structure constants, probabilities of electric dipole transitions, and atomic electric quadrupole moments. 
\hl{To check the 
predictive power of the
theory, we compared the respective characteristics with their experimental values available from previously published sources.}
Such comparisons are \st{of high importance} \hl{very important} for further developing theoretical approaches for ions with complex electronic structure, in particular those with an open $f$-shell. We also describe the experimentally observed Zeeman line splitting using semiempirical modeling.

The paper is structured as follows. In Section~\ref{sec:experiment} we describe the experimental technique and \st{intensity data processing} 
\hl{the method of processing the data}.
In Section~\ref{sec:theory} we present the theoretical approach and details of the calculations. Section~\ref{sec:results} contains the results. Finally, we conclude in Section~\ref{sec:conclusion}. 
Atomic units (a.u.) $\hbar = e = m_{e} = 1$ are used throughout the paper unless otherwise stated.
\section{Experiment} \label{sec:experiment}
In order to compare theoretical results for transition probabilities in \Tm\ with the experiment, we determined the relative intensity distributions in several branches of \Tm\ emission spectra starting from a common upper level. These spectra were recorded by a Bruker IFS-125HR Fourier transform spectrometer in the spectral range from 330 nm to 800 nm at spectral resolution 0.025 cm$^{-1}$ exploiting a quartz-vis beam splitter in the Laser Center of the University of Latvia~\cite{par_22}. To detect the radiation a Hamamatsu R928 photomultiplier tube with a wide spectral response was used. The excited Tm atoms were produced in a hollow cathode (HC) discharge lamp at two discharge currents, namely, 50 and 70~mA. 
The central part of the lamp was placed in a \st{big volume can} \hl{large vessel} (about 5 liters) filled with liquid nitrogen. The cathode was in the form of a hollow cylinder with an internal diameter of 3 mm and a length of 20 mm. The direction of light observation was collinear with the cathode.
At each current, the spectra were recorded with inert Ar and Ne gases. Thus, for the intensity analysis, four spectra were processed, which allowed us to minimize the statistical error of the intensity measurements.  The relative normalized intensity distributions were determined in each spectrum separately and then averaged. 

\begin{figure}
\includegraphics[width=0.9\columnwidth]{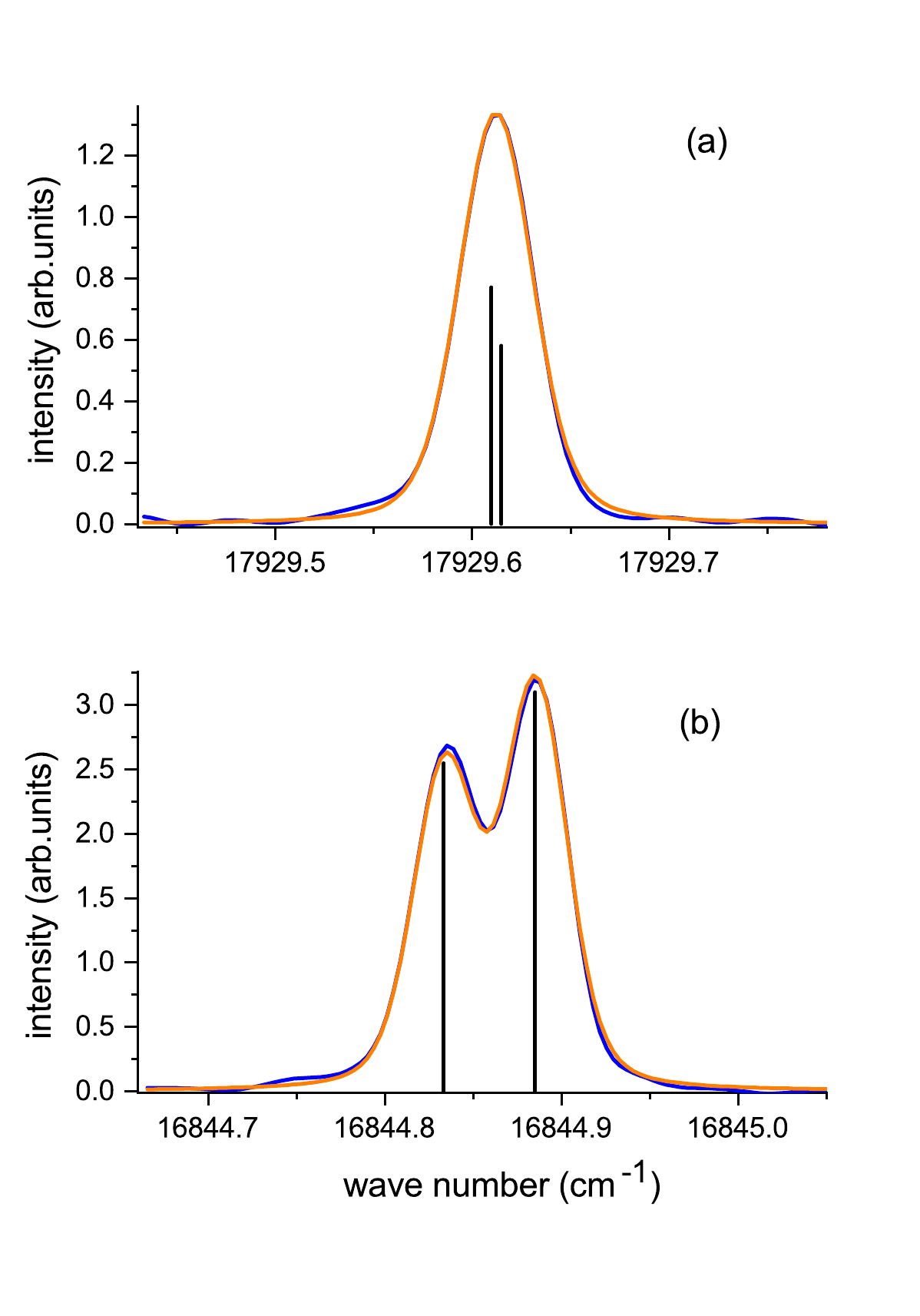}
\caption{Examples of intensity determination of the hyperfine components for the lines from the upper level with energy \num{35682.251}~cm$^{-1}$ \hl{($(4f^{12}5d6s6p)^o_{7/2}$)}: (a) line \num{17929.612}~cm$^{-1}$; (b) line \num{16844.863}~cm$^{-1}$; blue lines – experiment, orange lines – simulated line profile; vertical bars – respective intensity of hyperfine components.}
\label{fig:hf_int}
\end{figure}

For $^{169}$Tm the nuclear spin is I = $1/2$, hence the Tm lines consist of two strong and one or two, usually about 20 times weaker, hyperfine (hf) components. The observed hf splitting for most lines is small when compared to the line width determined by the Doppler broadening in the hollow cathode lamp and by the apparatus function of the FT spectrometer. Accordingly, for most of the analyzed lines the hf components were partially or completely overlapped, see Fig.~\ref{fig:hf_int}. 
The line intensity was represented as the sum of the two strongest hf components. Since we deal only with relative intensities in the branches with a common upper state level, excluding weaker hf components from intensity analysis introduces a negligible error. The intensity of hf components was determined by a fitting procedure to reproduce the recorded line profile. Each hf component was represented by a Voigt function with the frequency-dependent full width at half maximum being about \numrange{0.04}{0.05}~cm$^{-1}$. The intensity ratio of the two strongest hf components was kept fixed at the expected theoretical value, \hl{which depends} \st{dependent} on the total angular momenta of the levels.
As is seen from Fig.~\ref{fig:hf_int}, only intensities of the strongest components have been fitted.

Finally, the averaged intensity branching ratios were corrected according to the spectral sensitivity of the registration system, which consists of the FT spectrometer, including the beam-splitter, photomultiplier tube, output window of the HC lamp, and focusing lenses. 
\hl{To determine the spectral sensitivity function of our system in the visible range, we used as a reference the branching ratios of transition probabilities of Ar discharge emission lines from a common upper level reported in Ref.}~\cite{wha_93}. \hl{Overall, sixteen branches from Ref.}~\cite{wha_93}, \hl{containing 98 Ar lines, were selected. We have determined the relative intensity distributions for these branches in the spectra recorded in the present Tm experiment in an Ar-HC discharge, as well as in the Ar spectra recorded during our previous studies of other elements; see, for example, Ref.}~\cite{Ozdalgic_2019} \hl{for Ho and Ref.}~\cite{Guzelcimen_2014} \hl{for V. The obtained branching ratios were averaged over more than 20 such spectra thus diminishing the statistical error below 5\% for most of transitions. Note that in Ref.}~\cite{wha_93} \hl{the statistical error for 57 transitions does not exceed 5\%, while for 32 transitions the error spreads from 6\% to about 20\%.  Comparison of our ratio data with Ref.}~\cite{wha_93} has shown that in the range from about \num{16000}~cm$^{-1}$ to \num{25000}~cm$^{-1}$ \hl{the sensitivity is practically independent of frequency. However, some tiny diminishing of the sensitivity towards lower frequencies cannot be excluded. Below} \num{16000}~cm$^{-1}$ \hl{the sensitivity gradually diminishes, and for respective corrections in this range, we have used the spectral sensitivity curve from Ref.}~\cite{kli_12}.
%

Along with intensity measurements we have analyzed Zeeman splitting patterns of \Tm\ emission lines recorded in an external magnetic field. The same experimental setup was used for this purpose, except that a permanent magnet was tightly fixed above the cathode’s position just on top of the HC discharge tube. A cylindrical permanent magnet with a nominal strength of about 2000~G produced a vertical magnetic field. The distance between the magnet and the central part of the cathode emitting Tm radiation was about 15~mm. A linear polarizer was placed before the entrance of the FT spectrometer. The discharge current was 60~mA, and the Ar gas pressure was $1.3$~mbar. The spectra were recorded with a spectral resolution of 0.025~cm$^{-1}$ several times in the vertical ($\pi$-emission) and horizontal ($\sigma$-emission) direction of the polarization axes of the polarizer.
The magnetic field was calibrated using the splitting of the Ar lines, which is a common procedure in similar experiments, see, e.g., Ref.~\cite{Sobolewski_17}. In our case, we used the
Ar~\textsc{i} \num{14352.655}~cm$^{-1}$ and Ar~\textsc{ii} \num{21995.7839}~cm$^{-1}$ discharge lines. The calibration \st{led to the actual strength of the field, equal to 1820~G} \hl{determined that the actual strength of the field was 1820~G}.
%
\section{Theory} \label{sec:theory}
We \st{consider} \hl{treat} \Tm\ as a system with $15$ valence electrons and a frozen Xe-like core. For the computations of its atomic properties, we use the recently developed modification of the configuration interaction (CI) plus many-body perturbation theory (MBPT) approach~\cite{koz_22}. This variant of the original CI+MBPT method \cite{dzu_96} is designed for systems with a large number of valence electrons and uses different splitting of the many-electron space into the subspaces $P$ and $Q$, \st{where} \hl{in which} the non-perturbative and perturbative methods are used, respectively. It allows us to exclude double excitations to high-lying shells from the CI subspace $P$ and treat them perturbatively. The implementation of the CI+MBPT method within the corresponding computer package is presented in detail in Refs.~\cite{koz_15,che_21}. Here we only describe the \st{unique} aspects \hl{unique to} our present \st{consideration} \hl{work}.

We use a basis set $[11s10p9dfgh]$, where core (1-5$s$, 2-5$p$, 3-4$d$) and low-lying valence (6-7$s$, 6$p$, 5$d$, 4$f$) orbitals are obtained by solving the Dirac-Hartree-Fock equations in the $V^N$ potential, with $N$ being the total number of electrons. The higher-lying orbitals (8-11$s$, 7-10$p$, 6-9$d$, 5-9$f$, 5-9$g$, 6-9$h$) are generated using $B$ splines by the procedure described in Ref.~\cite{koz_19}.

The configuration subspace $P$ for the CI calculation is constructed in the following way. We keep the core shells frozen at all stages and make single (S) and double (D) excitations from several reference configurations to the valence orbitals 6-8$s$, 6-8$p$, 5-7$d$ and 4-6$f$.
We have chosen six odd and six even reference reference configurations: (odd) $4f^{13} 6s^2$, $4f^{13} 6s 5d$, $4f^{12} 6s^2 6p$, $4f^{12} 6s 5d 6p$, $4f^{13} 6s 7s$, and $4f^{12} 6s 6p 6d$; (even) $4f^{12} 6s^2 5d$, $4f^{13} 6s 6p$, $4f^{12} 6s^2 6d$, $4f^{13} 6s 7p$, $4f^{13} 5d 6p$, and $4f^{12} 5d 6p^2$. In the pure CI calculation we solve the matrix eigenvalue equation:
\begin{align} \label{Eq:CI}
    &  \hat{P}H\hat{P}\Psi_a= E_a \hat{P}\Psi_a\,,
\end{align}
where $\hat{P}$ is the projector on the $P$ subspace, $\Psi_a$ and $E_a$ are, \st{correspondingly} \hl{respectively}, the wave function and energy of the state $a$.
The Hamiltonian $H$ includes the magnetic part of the Breit interaction~\cite{koz_15}. We neglect the retardation and the quantum electrodynamics corrections since their contribution falls below our anticipated level of accuracy.

The subspace $Q$ complementary to subspace $P$ includes S and D excitations to the virtual orbitals 9-11$s$, 9-10$p$, 8-9$d$, 7-9$f$, 5-9$g$, 6-9$h$. In the CI+MBPT calculation, we follow the recipe from Ref.~\cite{koz_22} and split the subspace $Q$ into two parts, $Q_S$ and $Q_D$.
The subspace $Q_S$ is small in comparison to $P$, and we can redefine the subspace $P$, $P \to P'=P+Q_S$. The $Q_D$ subspace formed by D excitations from the valence to the virtual orbitals is taken into account using MBPT. We calculate effective potential $V_\mathrm{eff}(E)$ for the valence orbitals, which accounts for these excitations, and add it to the CI Hamiltonian $H$ when solving the matrix equation in the subspace $P'$:
\begin{subequations}
\label{Eq:CI+MBPT}    
\begin{align} \label{Eq:Heff}
    & H_\mathrm{eff}(E)=H+V_\mathrm{eff}(E)\,,    \\  
    \label{Eq:eigen}
    &  \hat{P'}H_\mathrm{eff}(E_a)\hat{P'}\Psi_a= E_a \hat{P'}\Psi_a\,,
\end{align}
\end{subequations}
where $\hat{P'}$ is the projector on the $P'$ subspace.
The energy dependence of the potential $V_\mathrm{eff}(E)$ is relatively weak, so we can use the same effective Hamiltonian for several eigenfunctions with close energies.

%
%
%

The typical sizes of the $P$ and $P'$ spaces in our calculations were $ \sim \numrange{9}{12}\times 10^3 $ relativistic configurations, which corresponds to $\sim \numrange{20}{25} \times 10^6$ Slater determinants.

\section{Results} \label{sec:results}
Our main goal is comparing newly measured relative intensity distributions in several branches of \Tm\ emission spectra originating from a common upper level with theoretical predictions. To achieve this, we first calculate the initial and final states of the desired transitions in order to determine their energies and wave functions. In calculations for a particular state, we also obtain all the low-lying states with the same parity $P$ and total angular momentum $J$. As a result, the total number of calculated levels in this work is substantial. In Section~\ref{subsec:spectra} we present only a subset of computed spectra to demonstrate the accuracy of our calculations. In Section~\ref{subsec:hyperfine} we also provide a comparison between the computed hyperfine structure constants and the experimental values for several levels, which tests the accuracy of the obtained wave functions at short distances. This is complementary to the comparison of the electric dipole transition intensities presented in Section~\ref{subsec:E1}, which serves as a test at large distances. Finally, in Section~\ref{subsec:EQM} we briefly discuss atomic electric quadrupole moments of the ground configuration levels, and in Section~\ref{subsec:magnetic} we present the measurements and modeling of line splitting in an external magnetic field.

\begin{figure}[hb]
\includegraphics[width=1.0\columnwidth]{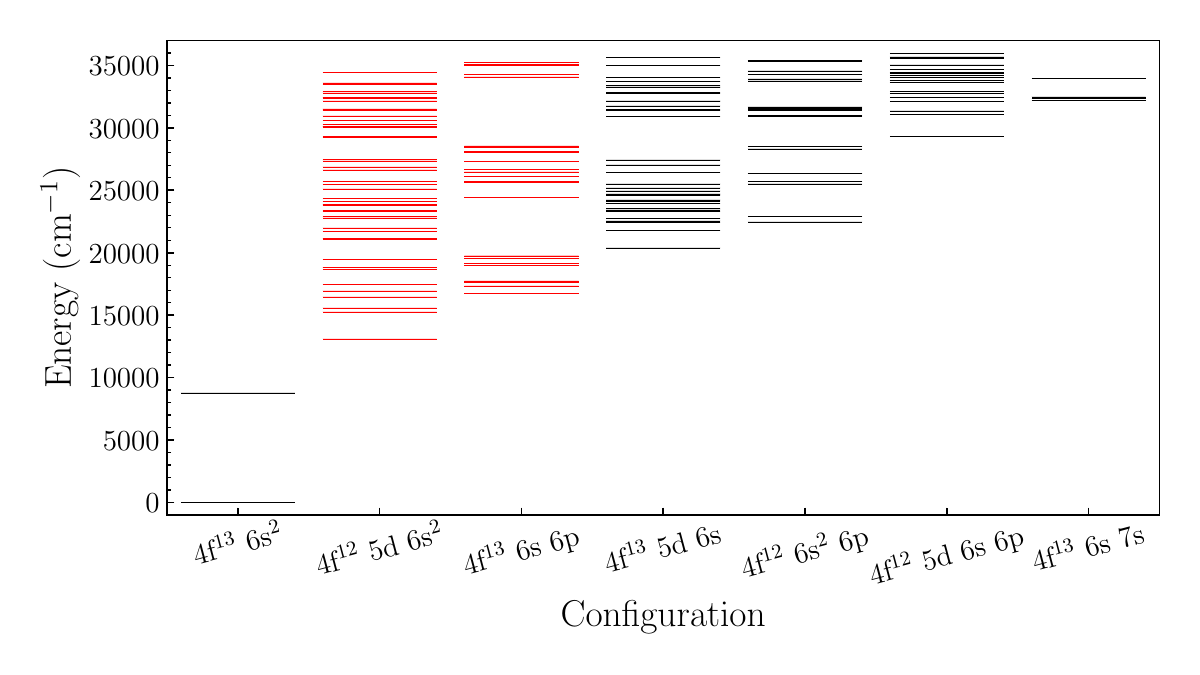}
\caption{Odd- (black) and even-parity (red) levels in \Tm\ with energies up to \num{36000}~cm$^{-1}$ according to the NIST ASD~\cite{NIST}.}
\label{fig:tm_nist_levels}
\end{figure}

\hl{To facilitate presenting the results, a scheme of the relevant subset of} \Tm\ \hl{levels is introduced in Fig.}~\ref{fig:tm_nist_levels}. \hl{The spectrum becomes considerably dense beyond} \num{15000}~cm$^{-1}$. \hl{Thus, the highest level considered in this study, with an energy of} \num{35682.251}~cm$^{-1}$, \hl{stands as the 161th level in the NIST ASD}~\cite{NIST}.
%
%
\subsection{Energies and $g$ factors} \label{subsec:spectra}
Let us start by presenting the calculated energies and and Land\'e $g$ factors for states of odd parity with a total angular momentum $J=7/2$ (as in the ground state, \hl{$(4f^{13}(^2F^o_{7/2})6s^2 )^o_{7/2}$, for example}). In Table~\ref{tab:spectrum_odd} we compare the results of our pure CI calculation (see Eq.~\eqref{Eq:CI}) with the experimental values from the NIST ASD~\cite{NIST}. 
\begin{table*}[]
    \caption{Low-lying energy levels of odd parity with $J=7/2$ in \Tm. The calculated energies (in cm$^{-1}$) and Land\'e $g$ factors are compared to the experimental values from the NIST ASD~\cite{NIST}.}
    \label{tab:spectrum_odd}
    \centering
    \begin{tabular}{cc|cc|ccc}
    \hline \hline
    \multicolumn{2}{c|}{} & \multicolumn{2}{c|}{Experiment\hl{~}\cite{NIST}} & \multicolumn{3}{c}{Theory} \\
     \cline{3-4} 
     \cline{5-7}
    Configuration & Term & Energy  & $g$ factor & Energy  &  Rel. diff. &  $g$ factor \\
    \hline
$ 4f^{13}(^2F^o)6s^2                  $ & $ ^2F^o      $ &     0.000  &  1.14119 &  0      &          &  1.14   \\ 
$ 4f^{13}(^2F^o_{7/2})5d6s(^3D)       $ & $ ^3[5/2]^o  $ & 23335.111  &  1.362   &  23275  & $ <1\% $ &  1.39   \\ 
$ 4f^{13}(^2F^o_{7/2})5d6s(^3D)       $ & $ ^3[9/2]^o  $ & 24246.425  &  1.02    &  24312  & $ <1\% $ &  1.06   \\ 
$ 4f^{13}(^2F^o_{7/2})5d6s(^3D)       $ & $ ^3[7/2]^o  $ & 24708.041  &  1.08    &  24629  & $ <1\% $ &  1.02   \\ 
$ 4f^{13}(^2F^o_{7/2})5d6s(^1D)       $ & $ ^1[7/2]^o  $ & 27037.468  &  1.11    &  26927  & $ <1\% $ &  1.12   \\ 
$ 4f^{12}(^3F_{4})6s^26p_{1/2}        $ & $ (4,1/2)^o  $ & 28340.290  &          &  29849  & $ -5\% $ &  1.17   \\ 
$ 4f^{13}(^2F^o_{5/2})5d6s(^3D)       $ & $ ^3[9/2]^o  $ & 30921.580  &  0.69    &  30972  & $ <1\% $ &  0.69   \\ 
$ 4f^{12}(^3H_{6})5d6s6p(^4F^o_{5/2}) $ & $ (6,5/2)^o  $ & 31367.728  &  1.33    &  31999  & $ -2\% $ &  1.39   \\ 
$ 4f^{12}(^3F_{4})6s^26p_{3/2}        $ & $ (4,3/2)^o  $ & 31694.749  &  1.17    &  33272  & $ -5\% $ &  1.12   \\ 
$ 4f^{13}(^2F^o_{7/2})6s7s(^3S_{1})   $ & $ (7/2,1)^o  $ & 32359.372  &  1.195   &  32263  & $ <1\% $ &  1.21   \\ 
$ 4f^{13}(^2F^o_{5/2})5d6s(^3D)       $ & $ ^3[5/2]^o  $ & 33240.362  &  1.17    &  33187  & $ <1\% $ &  1.21   \\ 
$ 4f^{13}(^2F^o_{5/2})5d6s(^3D)       $ & $ ^3[7/2]^o  $ & 33395.984  &          &  33408  & $ <1\% $ &  0.98   \\ 
$ 4f^{12}(^3H_{5})6s^26p_{3/2}        $ & $ (5,3/2)^o  $ & 33778.360  &          &  34777  & $ -3\% $ &  0.98   \\ 
$ 4f^{13}(^2F^o_{7/2})6s7s(^1S_{0})   $ & $ (7/2,0)^o  $ & 33961.044  &  1.15    &  34149  & $ -1\% $ &  1.14   \\ 
    \hline \hline
    \end{tabular}
\end{table*}
Remarkably, despite the challenges posed by calculations of atomic properties for higher-lying states, as presented in the table, we achieve good agreement with the experimental data for these levels, including those of the $4f^{13}(^2F^o_{7/2})6s7s$ configuration. The agreement for all the listed levels of configurations without excitations of $f$-electrons is consistently below 1\%. 
Excitations of the $f$-electrons lead to a more complex rearrangement of the electronic shell, causing our calculated energies for the $ 4f^{12}6s^26p $ levels to be less precise compared to the others. Our predicted Land\'e $g$ factors are compared to the available experimental values~\cite{NIST}, where applicable. Given the good agreement for these levels, we assume a comparable accuracy of less than 5\% for other levels in cases where experimental values are unavailable. It is worth mentioning that we observed similar trends and accuracy for the odd levels with different total angular moment $J$ value, which we do not present here. Since the accuracy of the obtained results is satisfactory for all the odd levels of our interest, we restrict ourselves to the pure CI calculation for these levels.

The final states of the desired transitions, except those leading to the ground configuration doublet $^2F^o$, have even parity. In Table~\ref{tab:spectrum_even}, we present the results for the low-lying states of even parity. We provide a comparison of the energy levels and Land\'e $g$ factors obtained in the pure CI and CI+MBPT models with the experimental data from the NIST ASD~\cite{NIST} and recent theoretical predictions by Li and Dzuba~\cite{li_20}. For both models, the energy of the lowest state is normalized to the experimental value.
\begin{table*}[]
    \caption{Low-lying energy levels of even parity in \Tm. The calculated energies (in cm$^{-1}$) \st{in} \hl{based on the} two models and Land\'e $g$ factors are compared to the experimental values from the NIST ASD~\cite{NIST} and theoretical data of Ref.~\cite{li_20}.
    }
    \label{tab:spectrum_even}
    \centering
    \begin{tabular}{ccc|cc|cc|ccc|cc}
    \hline \hline
    \multicolumn{3}{c|}{} & \multicolumn{2}{c|}{Experiment\hl{~}\cite{NIST}} & \multicolumn{7}{c}{Theory} \\
    \cline{6-12} 
    \multicolumn{3}{c|}{} & \multicolumn{2}{c|}{} & \multicolumn{2}{c|}{CI} & \multicolumn{3}{c|}{CI+MBPT} & \multicolumn{2}{c}{Ref.~\cite{li_20}} \\
    Configuration & Term  &  J & Energy & $g$ factor &   Energy   &  Rel. diff. & Energy    &  Rel. diff. & $g$ factor &  Energy & $g$ factor \\
    \hline
$ 4f^{12}(^3H_6)5d_{3/2}6s^2        $ & $(6,3/2)$ & $9/2 $  & $13119.610$ &  $1.305  $ & $13120$ &  $      $  & $13120$ &         & $ 1.31 $  &  $12780$   &  $1.3027$  \\
$ 4f^{12}(^3H_6)5d_{3/2}6s^2        $ & $(6,3/2)$ & $15/2$  & $15271.002$ &  $1.08   $ & $14830$ &  $ 3\%  $  & $14752$ & $ 3\% $ & $ 1.08 $  &   		   &  		    \\
$ 4f^{12}(^3H_6)5d_{3/2}6s^2        $ & $(6,3/2)$ & $11/2$  & $15587.811$ &  $1.255  $ & $15521$ &  $ <1\% $  & $15416$ & $ 1\% $ & $ 1.25 $  &   		   &  		    \\
$ 4f^{12}(^3H_6)5d_{5/2}6s^2        $ & $(6,5/2)$ & $17/2$  & $16456.913$ &  $1.175  $ & $15527$ &  $ 6\%  $  & $15581$ & $ 5\% $ & $ 1.17 $  &   		   &  		    \\
$ 4f^{13}(^2F^o_{7/2})6s6p(^3P^o_0) $ & $(7/2,0)$ & $7/2 $  & $16742.237$ &  $1.325  $ & $12545$ &  $ 25\% $  & $17730$ & $-6\% $ & $ 1.34 $  &  $16927$   &  $1.3120$  \\
$ 4f^{12}(^3H_6)5d_{5/2}6s^2        $ & $(6,5/2)$ & $7/2 $  & $16957.006$ &  $1.1722 $ & $16702$ &  $ 2\%  $  & $16750$ & $ 1\% $ & $ 1.17 $  &  $15574$   &  $1.1612$  \\
$ 4f^{13}(^2F^o_{7/2})6s6p(^3P^o_1) $ & $(7/2,1)$ & $7/2 $  & $17343.374$ &  $1.02153$ & $13087$ &  $ 25\% $  & $18262$ & $-5\% $ & $ 0.99 $  &  $17453$   &  $1.0235$  \\
$ 4f^{12}(^3H_6)5d_{3/2}6s^2        $ & $(6,3/2)$ & $13/2$  & $17454.818$ &  $1.15   $ & $17192$ &  $ 2\%  $  & $17027$ & $ 2\% $ & $ 1.13 $  &    		   & 		    \\
$ 4f^{13}(^2F^o_{7/2})6s6p(^3P^o_1) $ & $(7/2,1)$ & $9/2 $  & $17613.659$ &  $1.18598$ & $13409$ &  $ 24\% $  & $18606$ & $-6\% $ & $ 1.17 $  &  $17754$   &  $1.1877$  \\
$ 4f^{13}(^2F^o_{7/2})6s6p(^3P^o_1) $ & $(7/2,1)$ & $5/2 $  & $17752.634$ &  $1.186  $ & $13481$ &  $ 24\% $  & $18647$ & $-5\% $ & $ 1.21 $  &  $17867$   &  $1.1807$  \\
$ 4f^{12}(^3H_6)5d_{5/2}6s^2        $ & $(6,5/2)$ & $15/2$  & $18693.074$ &  $1.18   $ & $18063$ &  $ 3\%  $  & $17982$ & $ 4\% $ & $ 1.18 $  &    		   &  			\\
$ 4f^{12}(^3H_6)5d_{5/2}6s^2        $ & $(6,5/2)$ & $9/2 $  & $18837.385$ &  $1.1318 $ & $18092$ &  $ 4\%  $  & $18276$ & $ 3\% $ & $ 1.14 $  &  $16640$   &  $1.1277$  \\
$ 4f^{12}(^3H_6)5d_{5/2}6s^2        $ & $(6,5/2)$ & $11/2$  & $18853.823$ &  $1.15   $ & $18508$ &  $ 2\%  $  & $18311$ & $ 3\% $ & $ 1.10 $  &   		   &   		    \\
$ 4f^{13}(^2F^o_{7/2})6s6p(^3P^o_2) $ & $(7/2,2)$ & $11/2$  & $18990.406$ &  $1.215  $ & $14450$ &  $ 24\% $  & $19629$ & $-3\% $ & $ 1.27 $  &   		   &   		    \\
$ 4f^{13}(^2F^o_{7/2})6s6p(^3P^o_2) $ & $(7/2,2)$ & $3/2 $  & $19132.245$ &  $0.88   $ & $14558$ &  $ 24\% $  & $19738$ & $-3\% $ & $ 0.87 $  &  $19110$   &  $0.8699$  \\
$ 4f^{12}(^3H_6)5d_{5/2}6s^2        $ & $(6,5/2)$ & $13/2$  & $19466.663$ &  $1.15   $ & $18818$ &  $  3\% $  & $18676$ & $ 4\% $ & $ 1.15 $  &     	   & 			\\
$ 4f^{13}(^2F^o_{7/2})6s6p(^3P^o_2) $ & $(7/2,2)$ & $5/2 $  & $19548.834$ &  $0.983  $ & $15325$ &  $ 22\% $  & $20408$ & $-4\% $ & $ 0.96 $  &  $19485$   &  $1.0144$  \\
$ 4f^{13}(^2F^o_{7/2})6s6p(^3P^o_2) $ & $(7/2,2)$ & $9/2 $  & $19748.543$ &  $1.29   $ & $15433$ &  $ 22\% $  & $20521$ & $-4\% $ & $ 1.29 $  &    		   &		    \\
$ 4f^{13}(^2F^o_{7/2})6s6p(^3P^o_2) $ & $(7/2,2)$ & $7/2 $  & $19753.830$ &  $1.1839 $ & $15470$ &  $ 22\% $  & $20559$ & $-4\% $ & $ 1.18 $  &    		   &		    \\
    \hline \hline
    \end{tabular}
\end{table*}
\st{It is evident from Table~}\ref{tab:spectrum_even} \st{that here the excitations to the $Q$ subspace are crucial}. The low-energy part of the \Tm\ spectrum with even parity comprises levels belonging to either the $4f^{12} 5d 6s^2$ or $4f^{13} 6s 6p$ nonrelativistic configurations. In the simpler, pure CI calculation, which considers the $P$ subspace only, the levels of the latter configuration are too low, resulting in an approximately $25\%$ deviation from the experimental values. In the more advanced CI+MBPT calculation (see Eqs.~\eqref{Eq:CI+MBPT}), which accounts for both $P$ and $Q$ subspaces, these levels are elevated, leading to much better agreement with the experimental data. Meanwhile, the agreement for the levels of the other configuration remains largely unchanged. In fact, the account for the $Q$ subspace pushes the levels of the $4f^{13} 6s 6p$ configuration slightly above the experimental values, while the levels of the $4f^{12} 5d 6s^2$ configuration remain slightly below. \hl{Overall, the excitations to the $Q$ subspace, taken into account in the CI+MBPT, but not in the pure CI model, are crucial for even-parity levels}. The computed Land\'e $g$ factors exhibit only a minor dependence on the calculation model and are not shown for the pure CI calculation. It is worth noting that we display only the energy intervals for both models, even though their absolute energy values naturally differ. As the odd and even levels are calculated using different models, we apply additional normalization for presenting the energies of the even parity levels. This is not critical, as we ultimately use the experimental energies in calculating transition probabilities in Section~\ref{subsec:E1}.

Li and Dzuba \cite{li_20} recently calculated the electronic structure of \Tm\ using an approach that also combines configuration interaction with perturbation theory \cite{DBHF17}. 
They presented low-lying levels of even parity with $3/2 \leqslant J \leqslant 9/2$, which are connected by electric dipole transitions to the ground configuration doublet. As can be seen from the table, their results also show good agreement with the experimental data, especially for the levels of the $4f^{13} 6s 6p$ configuration. However, they did notice a discrepancy between their calculated $g$ factors and experimental values for a few states ~\cite{li_20}. It is revealed that the discrepancy arose because they inadvertently compared their levels \numrange{12}{14} (not shown here) to the NIST data for different levels. The correct comparison resolves the discrepancy.

\subsection{Hyperfine structure} \label{subsec:hyperfine}
Thulium has only one stable isotope with mass number $A=169$ and nuclear spin $I=1/2$. This leads to a relatively simple hyperfine structure, with no \hl{contribution from an} electric quadrupole moment \st{contribution}. It is characterized by the magnetic dipole hyperfine interaction constant $A_{\rm hfs}$, which has been measured for many levels (see Refs.~\cite{par_22,keb_22} for recent measurements). 
In Table~\ref{tab:hfs-even} we show the comparison between the \hl{presently computed and previously measured in Refs.}~\cite{Kuhl_71,Brandt_77,vLeeuwen_80,Childs_84,Pfeufer_86,Kroeger_97,par_22} values of $A_{\rm hfs}$ for the levels already presented in Table~\ref{tab:spectrum_even}. The order of the levels is the same as in Table~\ref{tab:spectrum_even}, so we omit their energies. The calculations are carried out in the CI+MBPT model. 
We adopted the value of the nuclear magnetic dipole moment of $^{169}$Tm from Ref.~\cite{Stone_05}, $\mu = -0.231\mu_N$, where $\mu_N$ is the nuclear magneton. 
\begin{table}
 \caption{
 Comparison of the \hl{presently calculated and previously measured} hyperfine structure constants $A_{\rm hfs}$ (in MHz) for low-lying levels of even parity in \Tm. 
 }
 \label{tab:hfs-even}
 \begin{lrbox}{\tablebox}
 \begin{tabular}{ccc|c|c}
 \hline \hline
Configuration & Term  &  J &  Experiment  & Theory  \\
 \hline
$ 4f^{12}(^3H_6)5d_{3/2}6s^2        $ & $(6,3/2)$  & $9/2 $  & $-441.5(12)^{\rm a}$  & $-484$  \\
$ 4f^{12}(^3H_6)5d_{3/2}6s^2        $ & $(6,3/2)$  & $15/2$  & $-345.18(23)^{\rm b}$  & $-334$  \\
$ 4f^{12}(^3H_6)5d_{3/2}6s^2        $ & $(6,3/2)$  & $11/2$  & $-390.46(28)^{\rm b}$  & $-413$  \\
$ 4f^{12}(^3H_6)5d_{5/2}6s^2        $ & $(6,5/2)$  & $17/2$  & $-308.89(17)^{\rm b}$  & $-307$  \\
$ 4f^{13}(^2F^o_{7/2})6s6p(^3P^o_0) $ & $(7/2,0)$  & $7/2 $  & $-736.6(10)^{\rm c}$  & $-653$  \\
$ 4f^{12}(^3H_6)5d_{5/2}6s^2        $ & $(6,5/2)$  & $7/2 $  & $-491.1(10)^{\rm d}$  & $-555$  \\
$ 4f^{13}(^2F^o_{7/2})6s6p(^3P^o_1) $ & $(7/2,1)$  & $7/2 $  & $-166.24(8) ^{\rm c}$  & $-205$  \\
$ 4f^{12}(^3H_6)5d_{3/2}6s^2        $ & $(6,3/2)$  & $13/2$  & $-365.91(55)^{\rm b}$  & $-368$  \\
$ 4f^{13}(^2F^o_{7/2})6s6p(^3P^o_1) $ & $(7/2,1)$  & $9/2 $  & $-629.25(8) ^{\rm c}$  & $-461$  \\
$ 4f^{13}(^2F^o_{7/2})6s6p(^3P^o_1) $ & $(7/2,1)$  & $5/2 $  & $-235.5(10)^{\rm d}$  & $-379$  \\
$ 4f^{12}(^3H_6)5d_{5/2}6s^2        $ & $(6,5/2)$  & $15/2$  & $-323.88(30)^{\rm b}$  & $-328$  \\
$ 4f^{12}(^3H_6)5d_{5/2}6s^2        $ & $(6,5/2)$  & $9/2 $  & $-422.4(9)  ^{\rm e}$  & $-492$  \\
$ 4f^{12}(^3H_6)5d_{5/2}6s^2        $ & $(6,5/2)$  & $11/2$  & $-476.51(27)^{\rm b}$  & $-378$  \\
$ 4f^{13}(^2F^o_{7/2})6s6p(^3P^o_2) $ & $(7/2,2)$  & $11/2$  & $-581.4(13)^{\rm f}$  & $-554$  \\
$ 4f^{13}(^2F^o_{7/2})6s6p(^3P^o_2) $ & $(7/2,2)$  & $3/2 $  &                        & $-8  $  \\
$ 4f^{12}(^3H_6)5d_{5/2}6s^2        $ & $(6,5/2)$  & $13/2$  & $-342.84(28)^{\rm b}$  & $-345$  \\
$ 4f^{13}(^2F^o_{7/2})6s6p(^3P^o_2) $ & $(7/2,2)$  & $5/2 $  & $-57(2)     ^{\rm g}$  & $-118$  \\
$ 4f^{13}(^2F^o_{7/2})6s6p(^3P^o_2) $ & $(7/2,2)$  & $9/2 $  & $-694.8(4)  ^{\rm f}$  & $-586$  \\
$ 4f^{13}(^2F^o_{7/2})6s6p(^3P^o_2) $ & $(7/2,2)$  & $7/2 $  & $-536.6(9)  ^{\rm g}$  & $-464$  \\
 \hline  \hline 
 \end{tabular}
  \end{lrbox}
\usebox{\tablebox}\\[1ex]
\parbox{\wd\tablebox}{
\raggedright
$^{\rm a}$ From Parlatan~\etal~\cite{par_22}.  \\
$^{\rm b}$ From Pfeufer~\cite{Pfeufer_86}. \\
$^{\rm c}$ From van~Leeuwen~\etal~\cite{vLeeuwen_80}. \\
$^{\rm d}$ From Childs~\etal~\cite{Childs_84}. \\
$^{\rm e}$ From Kuhl~\cite{Kuhl_71}. \\
$^{\rm f}$ From Kr\"oger~\etal~\cite{Kroeger_97}. \\
$^{\rm g}$ From Brandt and Camus~\cite{Brandt_77}.
}
 \end{table}
We observe that, for the majority of the levels, our \textit{ab initio} theoretical values for hyperfine structure constants are very close to the measurements, with the mean deviation of less than $60$~MHz. The absence of an experimental value for the $(4f^{13}(^2F^o_{7/2}) 6s6p(^3P^o_2))_{3/2}$ level can be explained by its very small value, which is less than $10$~MHz, according to our calculations. For this level, the dominant one-electron contributions to $A_{\rm hfs}$ coming from the $4f_{7/2}$ and $6s_{1/2}$ electrons, nearly cancel each other out.
Good agreement between the numerical results and the experimental data suggests that our calculated wave functions are sufficiently accurate.

The absence of stable thulium isotopes other than $^{169}$Tm is advantageous for our current experiment, as it eliminates the need for isotope separation.  However, the study of relative nuclear properties of isotopes, such as differences in root-mean-square charge radii, is hindered, because there is no way to extract the field-shift and mass-shift parameters with high precision from a King plot~\cite{King_84}. Therefore, accurate calculations of electronic structure and isotope shifts, which we will address in forthcoming work, are essential for studying properties of short-lived thulium isotopes.
\subsection{Intensities of dipole transitions} \label{subsec:E1}
We experimentally determine the branching ratios for several series of transitions that share a common upper level. For some of these transitions, the results are reported for the first time. For those already investigated in previous studies~\cite{wic_97,Wang_22}, we provide a comparison of the branching ratios. The statistical errors are derived from four measurements with a confidence limit of \num{0.95}. Additionally, we complement our experimental data with theoretical predictions. 

The probability (in s$^{-1}$ ) of an electric dipole transition from an upper state $u$ to a lower state $l$ is given by
\begin{equation}
A_{ul}= \frac{1}{2J_u+1}\frac{2.02613 \times 10^{18}}{\lambda^3} |\la \Psi_u || D || \Psi_l \ra|^2 ,
\end{equation}
where $J_u$ represents the total angular momentum of the upper state, $\lambda$ (in \AA) is the wavelength of the transition, and the reduced matrix element of the electric dipole operator $D$ is in atomic units.
The reduced matrix elements of the electric dipole operator, involving many-electron wave functions, are computed using the transition matrix approach (see Ref.~\cite{koz_15} for details). These wave functions are obtained from the CI or CI+MBPT models, as discussed earlier, and the experimental values are utilized for transition wavelengths. The reported values are computed in the length gauge. 
We expect that our calculated probabilities are reliable for sufficiently strong transitions with $A_{ul} \sim 10^6$~s$^{-1}$. 
Within the considered wavelength region this corresponds to reduced matrix elements on the order of 1 atomic unit. To assess the accuracy of the theoretical results more precisely, we rely on comparisons with experimental data. Conducting an independent, fully theoretical study to determine uncertainties in transition probabilities would be excessively complex in this case.

In Table~\ref{tab:prob_WL97_comp} we present our experimental and theoretical ratios of transition probabilities, $A_{ul}/A_{ul}^{\rm max}$, for lines originating from a common upper level, normalized to the strongest line in each series. Additionally, we provide our absolute theoretical values and compare them to previous measurements by Wickliffe and Lawler~\cite{wic_97}. 
\begin{table*}  
 \caption{Comparison of the experimental and calculated ratios of transition probabilities originating from a common upper level 
         \hl{and normalized to the strongest line in each series}.
         Our absolute theoretical values and previous results from Ref.~\cite{wic_97} are also shown. Energies and configurations of the levels are from the NIST ASD~\cite{NIST}.
  }
 \label{tab:prob_WL97_comp}
\begin{tabular}{T{0.25\textwidth}T{0.2\textwidth}T{0.13\textwidth}|T{0.08\textwidth}|T{0.06\textwidth}T{0.08\textwidth}|T{0.08\textwidth}T{0.12\textwidth}}
 \hline \hline
\multicolumn{1}{T{0.25\textwidth}}{Upper level (cm$^{-1}$)} & \multicolumn{1}{T{0.2\textwidth}}{Lower level (cm$^{-1}$)} & 
\multicolumn{1}{T{0.13\textwidth}|}{Transition wave number (cm$^{-1}$)} &
\multicolumn{3}{T{0.25\textwidth}|}{This work} & \multicolumn{2}{T{0.17\textwidth}}{Ref.~\cite{wic_97}} \\
\hline
\multicolumn{3}{T{0.58\textwidth}|}{} & \multicolumn{1}{T{0.08\textwidth}|}{Experiment} & \multicolumn{2}{T{0.14\textwidth}|}{Theory} & \multicolumn{2}{T{0.20\textwidth}}{Experiment} \\
&&& Ratio & Ratio &  $A_{ul}$  (s$^{-1}$) &  Ratio &   $A_{ul}$  (s$^{-1}$) \\
\cline{4-8}
31080.846 $(4f^{12}(^3H_6)5d6s6p(^4F^o_{3/2})^o_{11/2}$ & 13119.610 $(4f^{12}(^3H_6)5d_{3/2}6s^2)_{ 9/2}$ & 17961.236 & 1.000     & 1.000 & \cellcolor{bckgr} \num{7.08E+05} & 1.000     & \num{1.38(7)E+06}   \\ 
                                                        & 15587.811 $(4f^{12}(^3H_6)5d_{3/2}6s^2)_{11/2}$ & 15493.035 & not observed & 0.073 & \num{5.20E+04} & 0.047(4)  & \num{6.50(52)E+04}  \\ 
                                                        & 17454.818 $(4f^{12}(^3H_6)5d_{3/2}6s^2)_{13/2}$ & 13626.028 & 0.050(5)  & 0.073 & \num{5.16E+04} & 0.047(5)  & \num{6.50(59)E+04}  \\ \hline
32761.54 $(4f^{12}(^3H_6)5d6s6p(^4F^o_{5/2})^o_{17/2}$  & 15271.002 $(4f^{12}(^3H_6)5d_{3/2}6s^2)_{15/2}$ & 17490.538 & 1.000     & 1.00  & \num{4.53E+05} & 1.000     & \num{9.16(46)E+05}  \\ 
                                                        & 16456.913 $(4f^{12}(^3H_6)5d_{5/2}6s^2)_{17/2}$ & 16304.627 & 0.632(13) & 0.726 & \num{3.29E+05} & 0.612(43) & \num{5.61(28)E+05}  \\ 
                                                        & 18693.074 $(4f^{12}(^3H_6)5d_{5/2}6s^2)_{15/2}$ & 14068.466 & 0.117(7)  & 0.119 & \num{5.40E+04} & 0.105(13) & \num{9.60(106)E+04} \\ \hline
35633.01  $(4f^{12}(^3H_6)5d6s6p(^2D^o_{5/2})^o_{17/2}$ & 15271.002 $(4f^{12}(^3H_6)5d_{3/2}6s^2)_{15/2}$ & 20362.008 & 0.282(3)  & 0.263 & \num{1.68E+06} & 0.289(20) & \num{1.64(8)E+06}   \\ 
                                                        & 16456.913 $(4f^{12}(^3H_6)5d_{5/2}6s^2)_{17/2}$ & 19176.097 & 1.000     & 1.000 & \num{6.39E+06} & 1.000     & \num{5.67(28)E+06}  \\ 
                                                        & 18693.074 $(4f^{12}(^3H_6)5d_{5/2}6s^2)_{15/2}$ & 16939.936 & 0.331(10) & 0.271 & \num{1.73E+06} & 0.270(24) & \num{1.53(11)E+06}  \\ \hline
 \hline 
 \end{tabular}
 \end{table*}
The table reveals that our calculated values for the branching ratios are in very good agreement with both the current measurements and the experimental results of Ref.~\cite{wic_97}. 
However, the agreement for the absolute values is less satisfactory and diminishes with decreasing transition strength. The calculated probabilities for relatively weak lines originating from the level at \num{32761.54}~cm$^{-1}$ differ from the experimental values by nearly a factor of 2. On the other hand, for stronger transitions, the agreement is notably better. In fact, for the strongest lines originating from the level at \num{35633.01}~cm$^{-1}$ our results are almost consistent with the experimental uncertainties. 
In our measurements, the line at \num{15493.035}~cm$^{-1}$ was not observed, despite both theoretical predictions and previous experimental data suggesting that it has the same strength as the line at \num{13626.028}~cm$^{-1}$.

Next, in Table~\ref{tab:prob_gs} we present the results for transition probabilities to the ground state multiplet $^2F^o$. 
\begin{table*}  
 \caption{Comparison of the experimental and calculated ratios of transition probabilities originating from a common upper level to the ground configuration doublet 
         \hl{and normalized to the strongest line in each series}.
         Our absolute theoretical values and previous results from Refs.~\cite{wic_97,Wang_22} are also shown. Energies and configurations of the levels are from the NIST ASD~\cite{NIST}.
 }
 \label{tab:prob_gs}
\begin{lrbox}{\tablebox}
\begin{tabular}{T{0.25\textwidth}T{0.2\textwidth}T{0.13\textwidth}|T{0.08\textwidth}|T{0.06\textwidth}T{0.08\textwidth}|T{0.08\textwidth}T{0.13\textwidth}}
 \hline \hline
\multicolumn{1}{T{0.25\textwidth}}{Upper level (cm$^{-1}$)} & \multicolumn{1}{T{0.2\textwidth}}{Lower level (cm$^{-1}$)} & 
\multicolumn{1}{T{0.13\textwidth}|}{Transition wave number (cm$^{-1}$)} &
\multicolumn{3}{T{0.25\textwidth}|}{This work} & \multicolumn{2}{T{0.17\textwidth}}{Others} \\
\hline
\multicolumn{3}{T{0.58\textwidth}|}{} & \multicolumn{1}{T{0.08\textwidth}|}{Experiment} & \multicolumn{2}{T{0.14\textwidth}|}{Theory} & \multicolumn{2}{T{0.21\textwidth}}{Experiment} \\
&&& Ratio & Ratio &  $A_{ul}$  (s$^{-1}$) &  Ratio &   $A_{ul}$  (s$^{-1}$) \\
\cline{4-8}
21161.401  $(4f^{12}(^3F_4)5d_{3/2}6s^2)_{5/2}$ & 0.0 \quad  $(4f^{13}(^2F^o_{7/2})6s^2 )^o_{7/2}$ & 21161.401 & 1.000    & 1.000 & \num{2.24E+05} & 1.000    & \num{4.21(21)E+05}$^{\rm a}$   \\ 
                                                & 8771.243   $(4f^{13}(^2F^o_{5/2})6s^2 )^o_{5/2}$ & 12390.158 & 0.022(3) & 0.098 & \num{2.19E+04} & 0.019(3) & \num{8.00(120)E+03}$^{\rm a}$  \\ \hline
22791.176  $(4f^{12}(^3H_5)5d_{3/2}6s^2)_{7/2}$ & 0.0 \quad  $(4f^{13}(^2F^o_{7/2})6s^2 )^o_{7/2}$ & 22791.176 & 1.000    & 1.000 & \num{3.54E+06} & 1.000    & \num{3.69(13)E+06}$^{\rm b}$   \\ 
                                                & 8771.243   $(4f^{13}(^2F^o_{5/2})6s^2 )^o_{5/2}$ & 14019.933 & 0.005(1) & 0.005 & \num{1.62E+04} & 0.001(1) & \num{5.3(30)E+03}$^{\rm b}$    \\ \hline
22929.717  $(4f^{12}(^3H_5)5d_{5/2}6s^2)_{5/2}$ & 0.0 \quad  $(4f^{13}(^2F^o_{7/2})6s^2 )^o_{7/2}$ & 22929.717 & 1.000    & 1.000 & \num{5.73E+06} & 1.000    & \num{1.15(6)E+07}$^{\rm b}$    \\ 
                                                & 8771.243   $(4f^{13}(^2F^o_{5/2})6s^2 )^o_{5/2}$ & 14158.474 & 0.009(1) & 0.008 & \num{4.68E+04} & 0.002(1) & \num{1.82(50)E+04}$^{\rm b}$   \\ \hline
25717.197  $(4f^{12}(^3H_5)5d_{5/2}6s^2)_{7/2}$ & 0.0 \quad  $(4f^{13}(^2F^o_{7/2})6s^2 )^o_{7/2}$ & 25717.197 &          & 1.000 & \num{2.66E+07} & 1.000    & \num{3.72(19)E+07}$^{\rm a}$   \\ 
                                                & 8771.243   $(4f^{13}(^2F^o_{5/2})6s^2 )^o_{5/2}$ & 16945.954 &          & 0.025 & \num{6.63E+05} & 0.024(3) & \num{8.96(108)E+05}$^{\rm a}$  \\ \hline

 \hline 
 \end{tabular}
 \end{lrbox}
\usebox{\tablebox}\\[1ex]
\parbox{\wd\tablebox}{
\raggedright
$^{\rm a}$ From  Wickliffe and Lawler~\cite{wic_97}.  \\
$^{\rm b}$ From Wang~\etal~\cite{Wang_22}.
}
 \end{table*}
These series consist of two lines each, with the line leading to the ground state $^2F_{7/2}^o$ being approximately two orders of magnitude stronger than the line to the first excited state $^2F_{5/2}^o$. The factor \st{coming from} \hl{due to} the differences in wavelengths of these lines is around 4, while the rest arises from the differences in the reduced matrix elements. We observe very good agreement between the presently measured and calculated ratios. It is worth noting that our ratios for the second and third series are slightly larger than those reported in Ref.~\cite{Wang_22} and closer to the results of our calculations. However, dealing with such small transition probabilities poses challenges for theory and increases its uncertainty.
In the final series, originating from the level at \num{25717.197}~cm$^{-1}$, the theoretical ratio closely matches the result reported in Ref.~\cite{wic_97}, while our experimental ratio suffers from reabsorption issues in the setup and is not presented.
The discrepancy between our theoretical absolute values and previous measurements reported in Refs.~\cite{wic_97,Wang_22} for the stronger lines does not exceed a factor of 2. For the line at \num{22791.176}~cm$^{-1}$, our result even falls within the uncertainty of the experimental value.

\begin{figure}
\includegraphics[width=1.0\columnwidth]{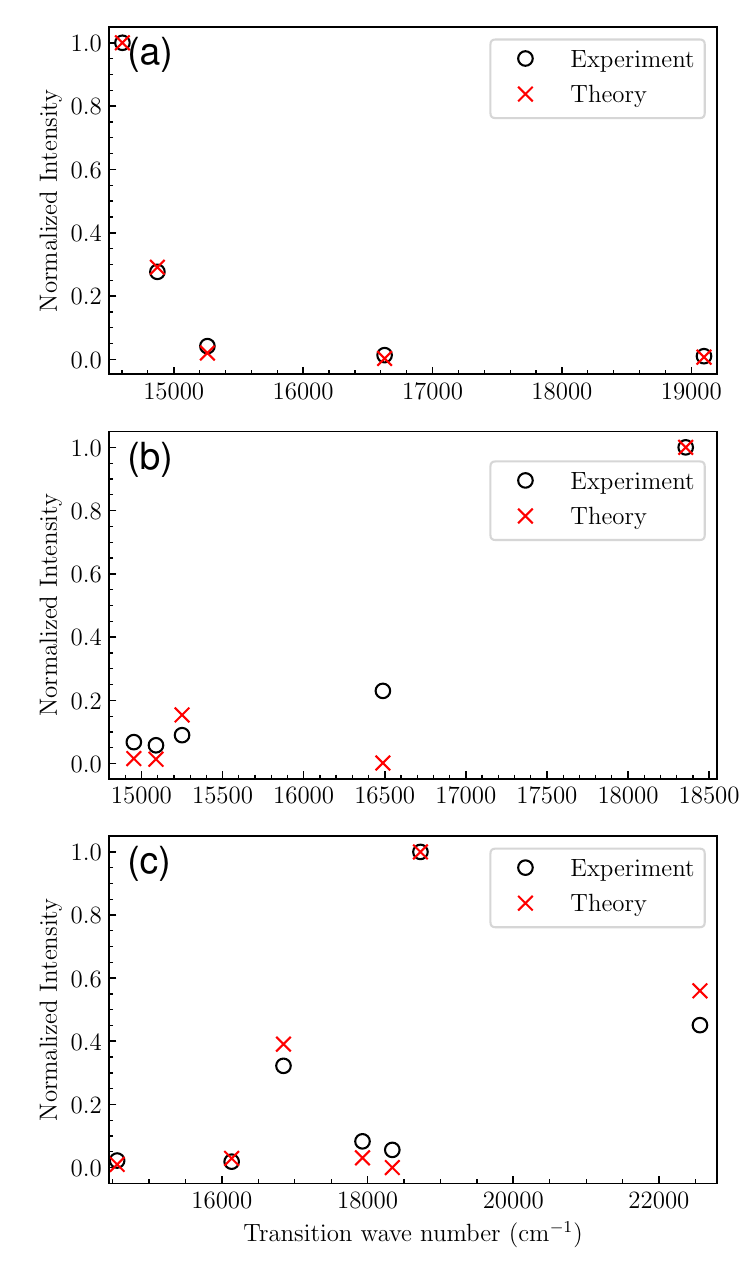}
\caption{Comparison of experimental and theoretical intensity distributions of emission lines originating from a common upper level at energies: (a)~\num{32217.195}~cm$^{-1}$, (b)~\num{33943.282}~cm$^{-1}$, (c)~\num{35682.251}~cm$^{-1}$.
}
\label{fig:Tmint}
\end{figure}

Finally, in Table~\ref{tab:prob_new} we present the comparison between our experimental and theoretical branching ratios for new series, that have not been published previously.  We also display our calculated absolute probabilities.
\begin{table*}  
 \caption{Comparison of the experimental and calculated ratios of transition probabilities from a common upper level
         \hl{and normalized to the strongest line in each series}.
         Our absolute theoretical values are also shown. Energies and configurations of the levels are from the NIST ASD~\cite{NIST}. 
 }
 \label{tab:prob_new}
\begin{tabular}{T{0.2\textwidth}T{0.25\textwidth}T{0.13\textwidth}|T{0.14\textwidth}|T{0.14\textwidth}T{0.14\textwidth}}
 \hline \hline
 \multicolumn{1}{T{0.2\textwidth}}{Upper level (cm$^{-1}$)} & \multicolumn{1}{T{0.25\textwidth}}{Lower level (cm$^{-1}$)} & \multicolumn{1}{T{0.13\textwidth}|} {Transition wave number (cm$^{-1}$)}  & \multicolumn{1}{T{0.14\textwidth}|}{Experiment} & \multicolumn{2}{T{0.28\textwidth}} {Theory} \\
 &&& Ratio & Ratio & $A_{ul}$  (s$^{-1}$) \\
 \hline
32217.195 $(4f^{13}(^2F^o_{7/2})6s7s(^3S_1))^o_{9/2}$ & 13119.610 $(4f^{12}(^3H_6)5d_{3/2}6s^2)_{ 9/2}$        & 19097.585 & 0.005(1) & 0.005 & \num{1.46E+04} \\
                                                      & 15587.811 $(4f^{12}(^3H_6)5d_{3/2}6s^2)_{11/2}$        & 16629.384 & 0.009(1) & 0.003 & \num{8.03E+03} \\
                                                      & 16957.006 $(4f^{12}(^3H_6)5d_{5/2}6s^2)_{7/2}        $ & 15260.189 & 0.037(1) & 0.019 & \num{5.15E+04} \\
                                                      & 17343.374 $(4f^{13}(^2F^o_{7/2})6s6p(^3P^o_1))_{7/2} $ & 14873.821 & 0.265(5) & 0.286 & \num{7.83E+05} \\
                                                      & 17613.659 $(4f^{13}(^2F^o_{7/2})6s6p(^3P^o_1))_{9/2} $ & 14603.536 & 1.000    & 1.000 & \num{2.74E+06} \\ \hline     
33943.282 $(4f^{12}(^3H_5)6s^26p_{3/2})^o_{13/2}$     & 15587.811 $(4f^{12}(^3H_6)5d_{3/2}6s^2)_{11/2}$        & 18355.471 & 1.000    & 1.000 & \num{3.94E+04} \\
                                                      & 17454.818 $(4f^{12}(^3H_6)5d_{3/2}6s^2)_{13/2}       $ & 16488.464 & 0.256(8) & 0.002 & \num{9.23E+01} \\
                                                      & 18693.074 $(4f^{12}(^3H_6)5d_{5/2}6s^2)_{15/2}$        & 15250.208 & 0.108(9) & 0.185 & \num{7.30E+03} \\
                                                      & 18853.823 $(4f^{12}(^3H_6)5d_{5/2}6s^2)_{11/2}       $ & 15089.459 & 0.070(6) & 0.018 & \num{6.93E+02} \\
                                                      & 18990.406 $(4f^{13}(^2F^o_{7/2})6s6p(^3P^o_2))_{11/2}$ & 14952.876 & 0.084(6) & 0.019 & \num{7.62E+02} \\ \hline
35682.251 $(4f^{12}5d6s6p)^o_{7/2}$                   & 13119.610 $(4f^{12}(^3H_6)5d_{3/2}6s^2)_{ 9/2}$        & 22562.641 & 0.374(8) & 0.465 & \num{1.17E+06} \\
                                                      & 16957.006 $(4f^{12}(^3H_6)5d_{5/2}6s^2)_{7/2}        $ & 18725.245 & 1.000    & 1.000 & \num{2.53E+06} \\
                                                      & 17343.374 $(4f^{13}(^2F^o_{7/2})6s6p(^3P^o_1))_{7/2} $ & 18338.877 & 0.057(2) & 0.001 & \num{5.25E+02} \\
                                                      & 17752.634 $(4f^{13}(^2F^o_{7/2})6s6p(^3P^o_1))_{5/2} $ & 17929.617 & 0.087(2) & 0.033 & \num{8.24E+04} \\
                                                      & 18837.385 $(4f^{12}(^3H_6)5d_{5/2}6s^2)_{9/2}        $ & 16844.866 & 0.358(14)& 0.434 & \num{1.10E+06} \\
                                                      & 19548.834 $(4f^{13}(^2F^o_{7/2})6s6p(^3P^o_2))_{5/2} $ & 16133.417 & 0.023(2) & 0.034 & \num{8.57E+04} \\
                                                      & 21120.836 $(4f^{12}(^3F_4)5d_{3/2}6s^2)_{7/2} $        & 14561.415 & 0.039(5) & 0.013 & \num{3.27E+04} \\ \hline
 \hline 
 \end{tabular}
 \end{table*}
The experimental and theoretical branching ratios are reasonably consistent with each other and align with the expectations of a simplified one-electron model. For example, consider the series with the upper level at \num{32217.195}~cm$^{-1}$. The strongest transition corresponds to an effectively one-electron $7s \to 6p$ transition, while conserving the total angular momentum $J$. Following that, the second strongest transition is also the same one-electron transition but with a change in $J$ from $9/2$ to $7/2$. The weaker transitions involve effectively two-electron processes: $(7s,4f) \to (6s,5d)$ transitions.
Similar considerations for the series originating from the level at \num{35682.251}~cm$^{-1}$ reveal that the lines induced by the $6p \to 6s$ one-electron transition are stronger than those resulting from the $5d \to 4f$ transition. In this case, the suppressed transition probability to the level at \num{21120.836}~cm$^{-1}$ can be attributed to an additional rearrangement involving the twelve $4f$ electrons, characterized by $^3F_4$ rather than $^3H_6$ quantum numbers. 
Lastly, all the lines starting from the level at \num{33943.282}~cm$^{-1}$ are weak. They are either produced by $6p \to 5d$ transitions accompanied by rearrangement of $4f$ electrons or by a $6s \to 4f$ transition with $\Delta l = 3$. Although our calculations generally capture the trend in this series, the transition probabilities are below our calculation capabilities.
We also offer a comparison of these branching ratios in Fig.~\ref{fig:Tmint} presented as normalized line intensity distributions.

It should be noted that in Table~\ref{tab:prob_new} the transition wave numbers, which are starting from the upper level 32217.195 cm$^{-1}$ \st{differ from the ones determined in our spectrum, being by about 0.04 cm$^{-1}$ lower} \hl{are about 0.04~cm$^{-1}$ lower than determined in our spectrum}. This difference might be \st{possibly} caused by relating these transition frequencies \st{rather} to the strongest hfs component (see, for instance, upper panel in Fig.~\ref{fig:Zeeman}) \st{and not} \hl{rather than} to the center of gravity.  Let us mention that for all other branches in Tables~\ref{tab:prob_WL97_comp} –  \ref{tab:prob_new} the transition frequencies determined from our spectrum are in agreement within 0.01 cm$^{-1}$ with the presented NIST values, as well as with the values in Ref.~\cite{wic_97}.

\subsection{Atomic electric quadrupole moments} \label{subsec:EQM}
After obtaining the wave functions of the desired many-electron states, we can calculate matrix elements corresponding to various atomic properties and processes. These include $g$ factors, hyperfine structure constants, amplitudes of multipole transitions, which have already been examined in this work, and others. Among them is the atomic electric quadrupole moment (EQM), which was recently calculated for the ground configuration doublet of \Tm\ by Fleig~\cite{fle_23}. The ground state EQM value $Q_{zz}(^2F_{7/2}) = 0.07^{+0.07}_{-0.00}$~a.u.
obtained in that paper through large-scale CI calculation is exceptionally small, making it favorable for atomic clock applications~\cite{Safronova_18}.
In this study, we corroborate this finding with our result of $0.06$~a.u. In contrast, a previous estimation obtained by Sukachev~\etal~\cite{Sukachev_16}, using the \textsc{cowan} code, yielded a value an order of magnitude larger, $Q_{zz}(^2F_{7/2}) \sim 0.5$~a.u.  For the EQM of the first excited state, we obtained a value of $Q_{zz}(^2F_{5/2}) = 0.05$~a.u., also in accordance with the value of Ref.~\cite{fle_23}. 
It is important to note that, for the EQM definition of an atom in an electronic state $|\gamma JM \ra$, having a total electronic angular momentum
$J$ and its projection $M$, we followed Refs.~\cite{fle_23,Angel_67}:
\begin{equation}
    Q_{zz} = -\frac{e}{2} \la \gamma JJ| \sum_i (3 z_i^2 - r_i^2) |\gamma JJ \ra,
\end{equation}
where the sum is over the atomic electrons and $\gamma$ stands for all other quantum numbers.
\subsection{Zeeman splitting} \label{subsec:magnetic}
The Zeeman splitting of the transition lines is an important source of information about the system. On the one hand, laboratory spectra in a well-calibrated magnetic field allow one to determine $g$ factors. This helps in assigning energy levels and determining their quantum numbers, especially in dense and complex spectra. When $g$ factors are already known, the splitting gives information about the magnetic field. This is particularly crucial in astrophysics, where it can be the only method to estimate the magnetic field.

\begin{figure}[htb]
\includegraphics[width=1.0\columnwidth]{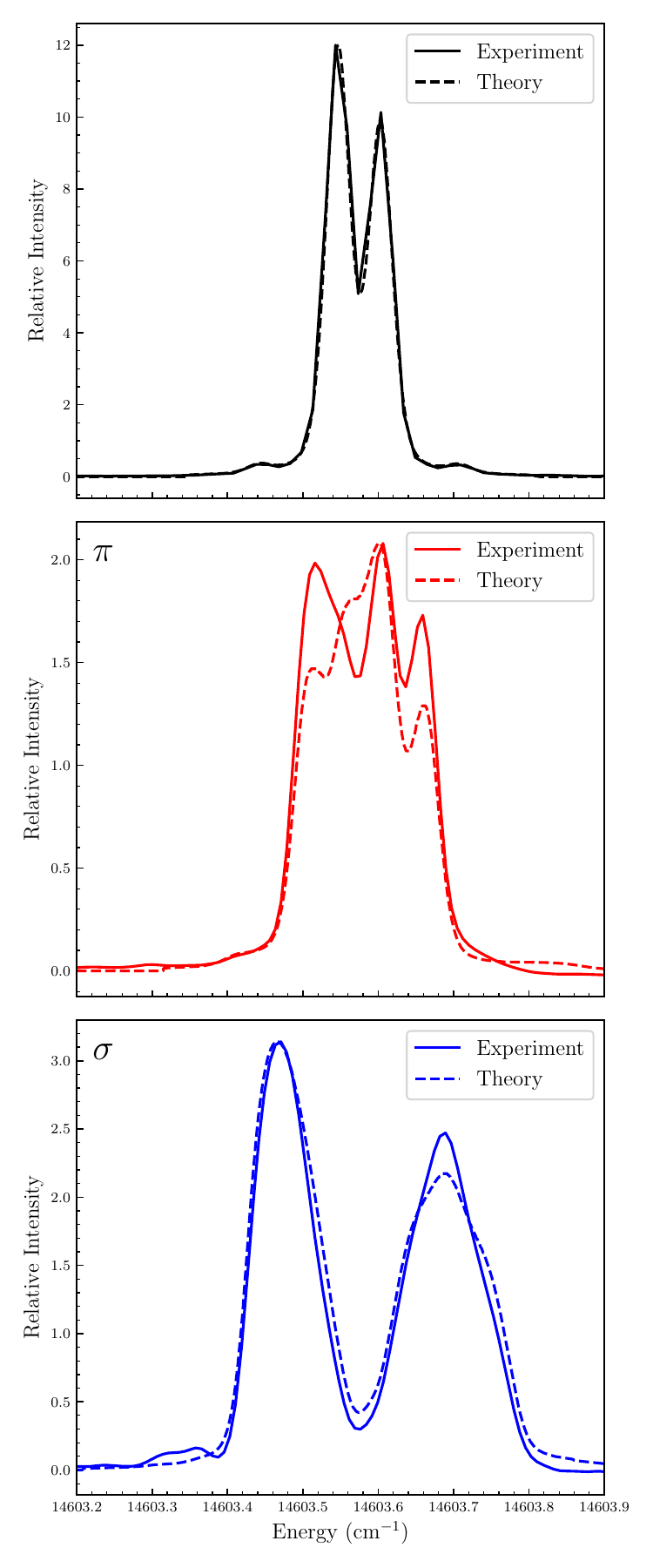}
\caption{Recorded transition between the levels $(4f^{13}(^2F^o_{7/2})6s7s(^3S_1))^o_{9/2}$ at \num{32217.195}~cm$^{-1}$ and $(4f^{13}(^2F^o_{7/2})6s6p(^3P^o_1))_{9/2}$ at \num{17613.659}~cm$^{-1}$. Top: without an external magnetic field; middle: $\pi$-component in the external magnetic field of 1820~G; bottom: $\sigma$-component in the external magnetic field of 1820~G. The amplitude of the experimental signal is normalized to the theory.
}
\label{fig:Zeeman}
\end{figure}

As an example, in Fig.~\ref{fig:Zeeman}, we present the experimentally recorded Zeeman splitting of the line corresponding to the transition between energy levels at \num{32217.195}~cm$^{-1}$ and  \num{17613.659}~cm$^{-1}$ (see Table~\ref{tab:prob_new}) in an external magnetic field of 1820~G.
In the theoretical analysis, we use the experimental values for $g$ factors from the NIST ASD~\cite{NIST} and hyperfine structure constants of the upper and lower levels from Refs.~\cite{Kroeger_97} and~\cite{vLeeuwen_80}, respectively.
One can see that such a magnetic field significantly alters the line shape. There is reasonably good agreement between theory and experiment. The difference can be partly attributed to misalignment between the polarizer and the magnetic field, leading to the mixing of $\sigma$ and $\pi$ polarizations.

Though the performed magnetic field experiments are rather of a descriptive kind, they \st{allow} \hl{make it possible} to demonstrate simultaneously the very different shapes of the Zeeman patterns for a huge number of lines in the high resolution FT spectrum. This might be useful for identifying the lines in dense spectra supplied by astrophysical observations when the presence of a magnetic field is expected. The recorded Zeeman splitting patterns of Tm lines are available upon request.
%
\section{Conclusion}  \label{sec:conclusion}
In this paper, we have presented a comparison between experimental and theoretical probabilities of electric dipole transitions in \Tm.
We measured intensities in several emission series originating from a common upper level. 
Since the population of the upper state is unknown, we derived relative transition probabilities within each series. The branching ratios for three of these series are presented for the first time. 
The experimental data are compared to the theoretical predictions obtained from large-scale calculations that combine configuration interaction with many-body perturbation theory.

Our findings reveal good agreement between the measurements and calculations. Furthermore, our predictions cover not only relative values but also absolute transition probabilities. When compared to the experimental absolute transition probabilities~\cite{wic_97,Wang_22}, our calculations demonstrate reliable accuracy, with a maximum deviation from the experiment of a factor of 2. For strong lines, the discrepancy is even smaller. Given the complexity of thulium's electronic structure, achieving this level of \textit{ab initio} theoretical precision can be considered quite good. 

Additionally, we provided predictions of the Land\'e $g$ factors and hyperfine structure constants of several levels \st{where} \hl{for which} experimental data are currently unavailable. Moreover, we corroborated recent theoretical results from Ref.~\cite{fle_23} regarding the atomic electric quadrupole moments of levels within the ground configuration doublet. Lastly, we showed a recorded example of the Zeeman splitting of a line in an external magnetic field compared to modeling and discussed a possible application of such spectra in astrophysics.

We want to emphasize that in the present study, we have achieved notably accurate theoretical results for transition probabilities originating from relatively high-lying levels in the open $f$-shell element. Such calculations are significantly more challenging than those involving low-lying levels only. We expect that our approach can yield valuable results for other ions with complex electronic structure.
%
\begin{acknowledgments}
This work has been started prior to February 2022.
We thank Marianna Safronova and Sergey Porsev for their calculations at the initial stage of the project.
The Riga team thanks Adams Lapins and Ilze Klincare for assistance in experiments, and Florian Gahbauer for helpful discussions. 
Support by the Latvian Council of Science, project No. lzp-2020/1-0088, and by the Scientific Research Projects Coordination Unit of Istanbul University, Project No. 30048, is acknowledged.
This research was supported in part through the use of DARWIN computing system: 
DARWIN -- A Resource for Computational and Data-intensive Research at the University of Delaware and in the Delaware Region, Rudolf Eigenmann, Benjamin E. Bagozzi, Arthi Jayaraman, William Totten, and Cathy H. Wu, University of Delaware, 2021, URL: https://udspace.udel.edu/handle/19716/29071. 
\end{acknowledgments}
\bibliography{thulium_biblio}{}
\bibliographystyle{h-physrev}
\end{document}